# Simulation analysis of the Compton-to-peak method for quantifying radiocesium deposition quantities

Malins A[1*], Ochi K[2], Machida M[1], and Sanada Y[2]

1. Center for Computation Science and e-Systems, Japan Atomic Energy Agency
2. Fukushima Environmental Safety Center, Japan Atomic Energy Agency

**Abstract**
Compton-to-peak analysis is one method for selecting suitable coefficients for converting count rates measured with in situ gamma ray spectrometry to radioactivity concentrations of $^{134}$Cs and $^{137}$Cs in the environment. The Compton-to-peak method is based on the count rate ratio of the spectral regions containing Compton scattered gamma rays to that with the primary $^{134}$Cs and $^{137}$Cs photopeaks. This is known as the Compton-to-peak ratio (RCP). RCP changes as a function of the vertical distribution of $^{134}$Cs and $^{137}$Cs within the ground. Inferring this distribution enables the selection of appropriate count rate to activity concentration conversion coefficients. In this study, the PHITS Monte Carlo radiation transport code was used to simulate the dependency of RCP on different vertical distributions of $^{134}$Cs and $^{137}$Cs within the ground. A model was created of a LaBr$_3$(Ce) detector used in drone helicopter aerial surveys in Fukushima Prefecture. The model was verified by comparing simulated gamma ray spectra to measurements from test sources. Simulations were performed for the infinite half-space geometry to calculate the dependency of RCP on the mass depth distribution (exponential or uniform) of $^{134}$Cs and $^{137}$Cs within the ground, and on the altitude of the detector above the ground. The calculations suggest that the sensitivity of the Compton-to-peak method is greatest for the initial period following nuclear fallout when $^{134}$Cs and $^{137}$Cs are located close to the ground surface, and for aerial surveys conducted at low altitudes. This is because the relative differences calculated between RCP with respect to changes in the mass depth distribution were largest for these two cases. Data on the measurement height above and on the $^{134}$Cs to $^{137}$Cs activity ratio is necessary for applying the Compton-to-peak method to determine the distribution and radioactivity concentration of $^{134}$Cs and $^{137}$Cs within the ground.

## 1. Introduction

In situ and airborne gamma ray spectrometry surveys are used for rapidly quantifying the amount of $^{134}$Cs and $^{137}$Cs at sites affected by radioactive fallout [1]. Information on the distribution of $^{134}$Cs and $^{137}$Cs with mass depth in the ground is essential for deriving activity concentrations from measured spectral count rates. A number of methods exist for analyzing gamma ray spectra to derive such information. These include comparing count rates of different energy peaks in the spectra [2], calculating the 'peak-to-valley' count rate ratio [3], and performing multiple in situ gamma ray spectrometry measurements using a collimator [4] or a lead plate [5].

The peak-to-valley method has historically been the most widely used method as it has the merits of (i) high sensitivity, (ii) being practicable for radionuclides with only a single intense gamma ray emission, and (iii) requiring only a single spectral measurement [6]. The principle is to calculate the ratio of count rates under a photopeak to those beneath the 'valley' region between the photopeak and its associated Compton edge. This ratio changes as a function of the mass depth distribution of the radioactivity within the ground. The deeper radionuclides are located within the ground, the greater the probability that the gamma rays they emit will undergo scattering interactions before reaching the detector, thus lowering the peak-to-valley ratio.

A method known as the Compton-to-peak method has been used for in situ gamma ray spectrometry surveys in Fukushima Prefecture [7, 8]. The method focusses on the ratio of count rates in the energy ranges containing the Compton continuum and the main photopeaks. This count rate ratio is also sensitive to the radioactivity distribution within the ground [7-9]. The merits of the Compton-to-peak method are (i) it alleviates issues relating to the close proximity of $^{134}$Cs and $^{137}$Cs photopeaks in gamma ray spectra [10], and (ii) it is applicable with low energy resolution gamma ray spectrometers, such as LaBr$_3$(Ce) and NaI(Tl) scintillators [7, 8].

This paper presents modelling of the characteristics of the Compton-to-peak method for estimating the vertical distribution and radioactivity concentrations of $^{134}$Cs and $^{137}$Cs within the ground. A model was created of a LaBr$_3$(Ce) detector used in drone helicopter aerial radiation surveys in Fukushima Prefecture [7] for Monte Carlo radiation transport simulations. The model was verified by comparing simulated spectra to measurements from test sources in a laboratory. The characteristics of the Compton-to-peak method were analyzed by calculating the Compton-to-peak ratio for different vertical distributions of $^{134}$Cs and $^{137}$Cs within the ground and for different altitudes for taking measurements.

## 2. Methods
### 2.1. Simulation Model

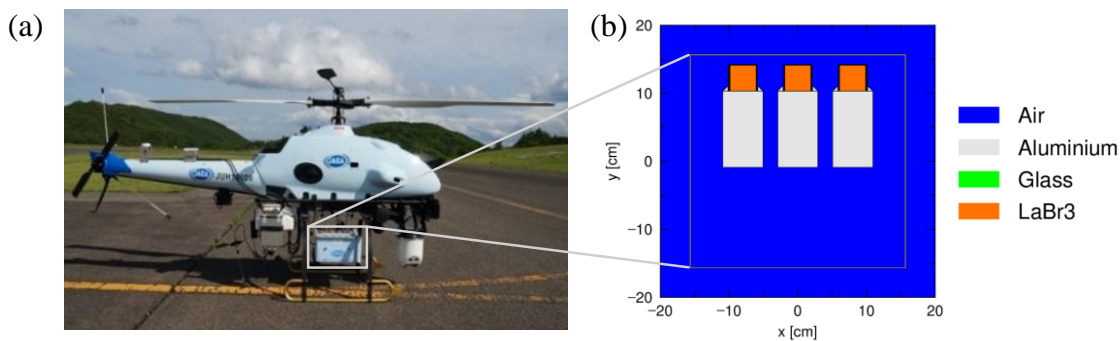

Figure 1. (a) LaBr$_3$(Ce) detector mounted underneath the drone helicopter used for airborne surveys. (b) Top-down view of the detector model created in PHITS.

A simulation model was created based on engineering drawings of the LaBr$_3$(Ce) detector used in drone helicopter surveys in Fukushima Prefecture. The model contained representations of the three LaBr$_3$(Ce) scintillation crystals, their aluminum and glass casings, the photomultiplier tubes and the outer housing of the detector (Fig. 1). The elemental compositions of the materials in the model followed references [11, 12]. The Particle and Heavy Ion Transport code System (PHITS, ver. 3.02) was used for Monte Carlo radiation transport simulations [13]. The EGS5 option in PHITS was turned on for modelling the atomic interactions of photons, electrons and positrons [14]. Gamma ray spectra were calculated by simulating the transport and interactions of gamma rays and charged particles, and tallying the energy deposited by charged particles within the LaBr$_3$(Ce) scintillation crystals. The tallies had 3 keV wide energy bins to match the output from the detector.

## 2.2. Energy Dependency of the Detector Resolution

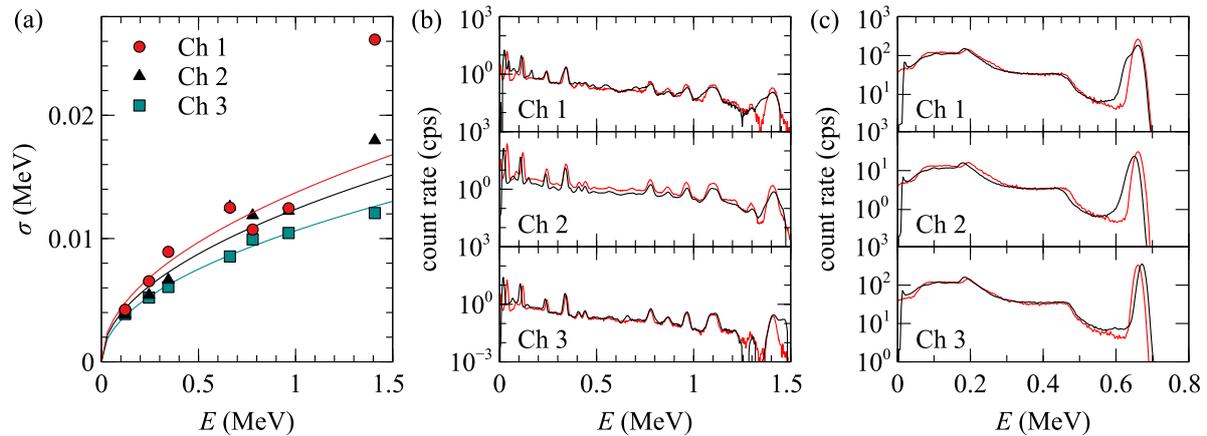

Figure 2. (a) Peak widths extracted from measured $^{152}$Eu and $^{137}$Cs spectra. Lines show fits to $\sigma = aE^{0.5}$. (b) Comparison of measured spectra after background subtraction (black lines) and simulated spectra (red lines) for a $^{152}$Eu source. (c) Like (b) but spectra for $^{137}$Cs source.

The energy dependency of the resolution of each LaBr$_3$(Ce) scintillation crystal was established from gamma ray spectra measured from test sources. Spectra were measured from a 7450 Bq $^{152}$Eu source taped to the side of the detector housing (adjacent to the central LaBr$_3$(Ce) scintillation crystal (channel 2)), and from a 10.6 MBq $^{137}$Cs source placed centrally 20 cm below the detector. Background count rates were subtracted from the measured spectra. The widths of peaks in the spectra associated with a single energy gamma ray emission were extracted using the Covell method [15]. The energy dependency of the peak widths was fitted using the function $\sigma = aE^{0.5}$ for each detector channel (Fig. 2(a)). The results were used to model the detector energy resolution in PHITS by fluctuating randomly the energies deposited by charged particles within each LaBr$_3$(Ce) scintillation crystal by a Gaussian distribution with standard deviation given by the respective $\sigma$ formula.

## 2.3. Simulation Cases

The simulation model was verified by comparing simulation results against spectra measured from the $^{152}$Eu and $^{137}$Cs test sources. A series of spectra were measured with the $^{137}$Cs source fixed with tape at the bottom of a plastic vessel and with the LaBr$_3$(Ce) detector suspended above the vessel on a hoist. The distance between the source and the bottom of the detector housing was varied between 20 and 50 cm. The level of water in the vessel was varied between 0 and 30 cm. Compton-to-peak ratios (RCP) were calculated from the spectra as

$$RCP = \Sigma C_{50\text{-}450\,keV} / \Sigma C_{450\text{-}760\,keV} \qquad (1)$$

where $C_{50\text{-}450\,keV}$ and $C_{450\text{-}760\,keV}$ are the count rates for energy ranges 50-450 keV and 450-760 keV, respectively. Note the sum is over the three LaBr$_3$(Ce) channels. Net count rates after background subtraction were used in the case of spectra measured in the laboratory.

The Compton-to-peak method for assessing radioactivity levels of $^{134}$Cs and $^{137}$Cs in the environment was analyzed by performing simulations of the detector in the infinite half-space geometry. This geometry consists of two infinite slabs of matter representing soil and air which are separated by a horizontal plane [16]. Simulations were performed with $^{134}$Cs and $^{137}$Cs (i) distributed exponentially with mass depth in soil (characterized by the relaxation mass depth parameter $\beta$), and (ii) distributed uniformly in the ground from the surface down to a mass depth of 40 g/cm$^2$. The height of the bottom of the detector above the ground surface was set at 10, 25 and 50 m in different simulations.

## 3. Results
### 3.1. Comparison of Simulations and Measurements

Fig. 2(b) and (c) respectively show simulated and measured spectra for the $^{152}$Eu source located on the side of the detector and for the $^{137}$Cs source located 20 cm below the detector. A good correspondence was obtained between both sets of spectra in terms of spectrum shape, peak positions and peak widths.

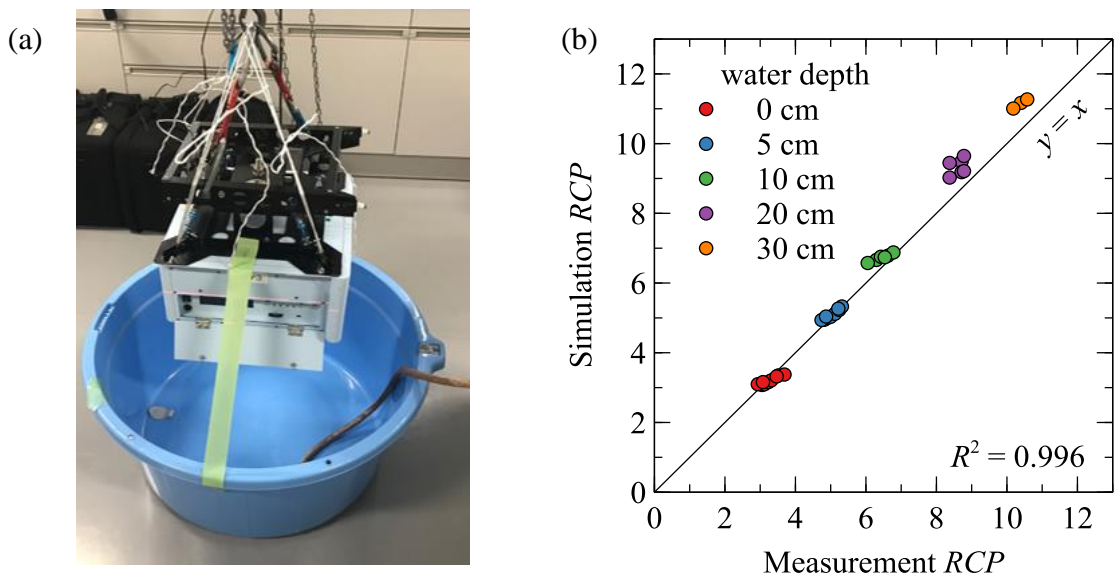

Figure 3. (a) Detector suspend above the $^{137}$Cs source located at the bottom of the plastic water vessel. (b) Correlation of simulated and measured RCP for different detector positions and water levels. Colored markers show different depths of water in the vessel. Markers with same color show results for different channels and distances between the source and the detector.

Fig. 3(a) shows the experimental apparatus used to test whether RCP values from simulated spectra were compatible with those from measured spectra. Increasing the height of the detector above the $^{137}$Cs source, or increasing the level of water in the plastic vessel, raises RCP as the amount of gamma ray scattering between the source and the detector increases. Fig. 3(b) shows there is a strong correlation between RCP values extracted from the simulated and measured spectra. As water is denser than air, increasing the water depth has a larger effect on RCP than increasing the distance between the source and the detector.

## 3.2. Simulation Analysis of the Compton-to-peak Method

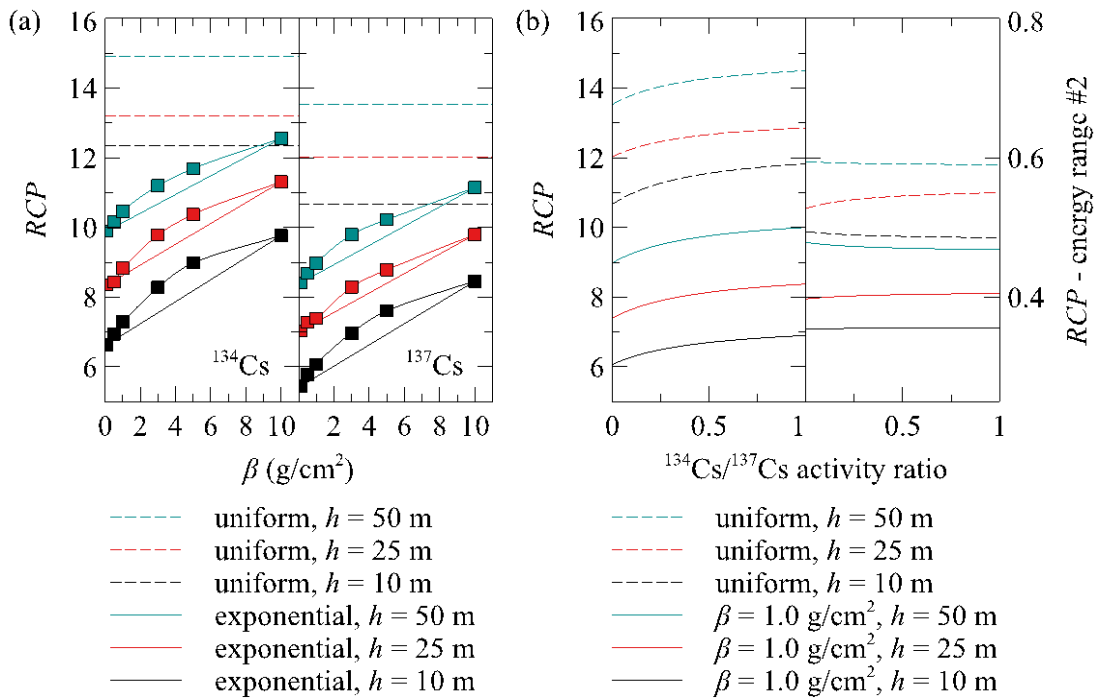

Figure 4. (a) Simulated RCP for different mass depth distributions of $^{134}$Cs (left panel) and $^{137}$Cs (right panel) within the ground. Solid lines show results for exponential distribution, and dashed lines results for the uniform distribution. Colors indicate different heights of the detector above the ground. (b) Simulated dependencies of RCP on the $^{134}$Cs to $^{137}$Cs activity ratio. Left panel shows results for RCP calculated via eq. (1). Right panel shows results for count rate ratio using energy ranges 447-552 keV and 552-849 keV.

The simulation results for the dependency of RCP on the mass depth distribution of $^{134}$Cs and $^{137}$Cs within the ground are shown in Fig. 4(a). RCP increases with the relaxation mass depth parameter $\beta$ for $^{134}$Cs and $^{137}$Cs with an exponential distribution within the ground. The rate of increase of $\beta$ decreases as the magnitude of $\beta$ increases. This means that the RCP signal will be less sensitive for differentiating between mass depth distributions with high values of $\beta$ than with low values of $\beta$. Note that low values of $\beta$ typically occur in the first years following radioactive fallout when radioactivity is located close to the ground surface [17].

The dashed lines in Fig. 4(a) show RCP results for $^{134}$Cs and $^{137}$Cs distributed uniformly within the ground down to 40 g/cm$^2$. There are clear separations between the RCP values for the uniform distributions and those for the exponential distribution of the radioactivity within the ground. The range of $\beta$ values covered in the latter case (0.1 to 10 g/cm$^2$) are typical for sites that have not been decontaminated nor significantly altered since the original fallout [17]. Given the large separation of RCP values between the uniform and exponential distribution cases, it is expected that the Compton-to-peak method will be useful for remotely sensing if sites have undergone mixing of the soil, or decontamination by interchanging soil layer positions, or by topsoil removal and recovering with clean soil [18]. These processes yield radioactivity distributions that are approximately uniform or deeper than uniform within the ground, leading to even higher RCP values in the latter case.

The different colored lines in Fig. 4(a) show the dependency of RCP on the altitude of the detector above the ground. The differences in RCP between different detection heights are comparable to the differences with varying $\beta$. Having an accurate estimate of the altitude at which a measurement was taken will thus be important for practical application of the Compton-to-peak method. As count rates are higher closer to the ground, and the relative differences in RCP between different values of $\beta$ are greater at lower heights, low altitude airborne surveys are expected to offer the greatest sensitivity for determining $^{134}$Cs and $^{137}$Cs mass depth distributions and activity concentrations.

The effects of land topography [19, 20], vegetation and other structures [21-23] on the Compton-to-peak method were not assessed in this study which only modelled the infinite half-space geometry. Further research is needed on how these factors affect the application of the Compton-to-peak method in practice. Nonetheless, there is still a range of sites for which these modelling results should be applicable, including agricultural land, paddy fields, parks, sports fields and brownfield sites.

The two panels in Fig. 4(a) demonstrate that RCP values for $^{134}$Cs and $^{137}$Cs are different given identical distributions for the radioactivity within the ground. Therefore the $^{134}$Cs to $^{137}$Cs activity ratio must also be taken into account when applying the Compton-to-peak method. The left panel of Fig. 4(b) shows how RCP changes as a function of the $^{134}$Cs to $^{137}$Cs activity ratio. In all cases RCP decreases as $^{137}$Cs becomes the predominant radionuclide.

By changing the energy ranges for calculating RCP it is possible to obtain results with lower dependency on the $^{134}$Cs to $^{137}$Cs activity ratio. The right panel of Fig. 4(b) shows one example. In this case RCP was calculated as the ratio of net count rates in the energy ranges 447-552 keV and 552-849 keV. These energy ranges give a ratio more alike to the Peak-to-valley method as they compare the counts below the main valley region between the Compton edge and the photopeaks to the main energy photopeaks. Note care will be needed with background subtraction to apply this energy range in practice, as the energy region above 760 keV is affected by self-contamination by radioactive $^{138}$La within the detector. This version of RCP is clearly less sensitive to the $^{134}$Cs to $^{137}$Cs activity ratio (cf. left and right panels of Fig. 4(b)). However it has the drawbacks of smaller relative separations between the different mass depth distribution and measurement height cases, and greater statistical uncertainty for a given counting time due to lower total count rates in the applicable energy ranges.

4.  Conclusions

A Monte Carlo radiation transport model was created of a LaBr$_3$(Ce) detector used for drone helicopter radiation surveys. The model was verified using spectra measured from radioactive $^{152}$Eu and $^{137}$Cs sources. The characteristics of the Compton-to-peak method for assessing $^{134}$Cs and $^{137}$Cs mass depth distributions and radioactivity concentrations in the environment were analyzed using the model. The simulations demonstrated that RCP depends on both measurement height above the ground and the $^{134}$Cs to $^{137}$Cs activity ratio. Both these factors must be accounted for when using RCP to infer the radioactivity distribution within the ground. Measurements taken near to the ground surface are expected to give the best results, as relative differences in RCP for different radioactivity distributions in the ground are greatest in this case. The method is expected to be useful for distinguishing between typical exponential distributions of radioactivity in the ground and distributions where the radioactivity is located deeper in the ground on average, e.g. due to decontamination or soil mixing. Once a mass depth distribution for $^{134}$Cs and $^{137}$Cs radioactivity within the ground has been inferred with the Compton-to-peak method, it is then possible to select conversion coefficients for net spectral count rates to calculate $^{134}$Cs and $^{137}$Cs radioactivity concentrations in the environment. Note the analysis in this paper assumes background count rates can be subtracted from measured spectra prior to performing Compton-to-peak analysis.

**Acknowledgements**
The simulations were performed on JAEA's SGI ICEX supercomputer. We thank colleagues in the Center for Computational Science & e-Systems and the Environmental Research Group of CLADS within JAEA for their support.